# Resolving the emission transition dipole moments of single doubly-excited seeded nanorods *via* heralded defocused imaging

*Daniel Amgar[1,†], Gur Lubin[1,†], Gaoling Yang[2], Freddy T. Rabouw[3], and Dan Oron[4,\*]*

[1] Department of Physics of Complex Systems, Weizmann Institute of Science, Rehovot 76100, Israel

[2] School of Optics and Photonics, Beijing Institute of Technology, China

[3] Debye Institute of Nanomaterials Science, Utrecht University, Princetonplein 1, 3584 CC Utrecht, The Netherlands

[4] Department of Molecular Chemistry and Materials Science, Weizmann Institute of Science, Rehovot 76100, Israel

ABSTRACT

Semiconductor nanocrystal emission polarization is a crucial probe of nanocrystal physics and an essential factor for nanocrystal-based technologies. While the transition dipole moment of the lowest excited state to ground state transition is well characterized, the dipole moment of higher multiexcitonic transitions is inaccessible *via* most spectroscopy techniques. Here, we realize direct characterization of the doubly-excited state relaxation transition dipole by heralded defocused imaging. Defocused imaging maps the dipole emission pattern onto a fast single-photon avalanche diode detector array, allowing the post-selection of photon pairs emitted from the biexciton–exciton emission cascade and resolving the differences in transition dipole moments. Type-I½ seeded nanorods exhibit higher anisotropy of the biexciton-to-exciton transition compared to the exciton-to-ground state transition. In contrast, type-II seeded nanorods display a reduction of biexciton emission anisotropy. These findings are rationalized in terms of an interplay between transient dynamics of the refractive index and the excitonic fine structure.

INTRODUCTION

In recent years, spectroscopy of single particles has evolved to provide profound understanding of intrinsic properties of semiconductor nanocrystals (NCs). This is due to a combination of advances in synthesis and of dramatic advances in single-photon detection systems. The improvements in synthesis protocols over the years yield photo-stable, tunable, and bright semiconductor nanocrystals (NCs), enabling deeper spectroscopic investigations at the single-particle level as well as utilization in plethora of applications. One of the most important features of NCs, as light emitters, is polarization of optical transitions, experimentally characterized by the dimensionality and orientation of their excitation and emission transition dipoles.[1] Emission polarization from CdSe spherical microcrystals, which originates from their hexagonal lattice, has been theoretically and experimentally reported in the early 1990s.[2,3] Nanorod (NR) architectures, quantum-confined in two dimensions, have been shown to preferentially emit linearly polarized light due to a combination of the anisotropic lattice and dielectric effects due to the anisotropic shape. According to previous reports, the highly-polarized NR emission indicates that the emissive transition has a one-dimensional (1D) dipole along the long axis of the rods.[4–6] Several works have used polarization microscopy to determine the three-dimensional (3D) orientation of single symmetric fluorophores, with the requirement to have twofold degenerate transition dipole oriented isotropically in two dimensions.[7–10] Alternatively, Jasny *et al.* and other groups have proposed the method of defocused imaging for finding the 3D orientation of a radiating dipole based on the analysis of light intensity distribution, supported by theoretical calculations.[11–16] Notably, however, all the above works have characterized the transition dipole moment of the singly-excited state relaxation, which typically accounts for most of the emission in photoluminescence experiments. The transition dipole moments of higher (multiply) excited states remain largely unexplored. Indeed, spectroscopic characterization of multiply-excited states is difficult due to their low emission quantum yield and rapid decay dynamics.[17] Yet, since multiply-excited states play a crucial role for achieving optical gain and in the development of quantum light sources, developing a sophisticated, yet facile, approach to directly probe the dipole moment of higher-excited states at the single-particle level is required. Specifically, much attention has been directed towards spectroscopic characterization of the doubly-excited or biexciton (BX) state.[18,19] A BX is formed upon double excitation of a single NC and may decay to the ground state (GS) through a cascaded process, emitting two subsequent photons. The first one can be assigned to the BX-to-exciton (X) transition and the second one to the X-to-GS transition, termed here BX and X photons, respectively. Recently, Heralded Spectroscopy has demonstrated unambiguous isolation of multiexciton emission through temporal photon correlations using single-photon avalanche diode (SPAD) array detectors.[20,21] Isolating the BX state from other emissive states through photon correlations features significant advantages over ensemble detection schemes, such as transient absorption (TA) and time-resolved photoluminescence (PL), in which the interpretation of the results is more challenging due to a combination of inhomogeneous broadening and contributions from higher excited, often also charged, states.[22–24] Crucially for this work, the heralded post-selection of BX and X emission is a single-NC method rather than an ensemble technique and as such does not average over emitter orientations. Combining defocused imaging with heralded post-selection of multiexciton emission can provide an elegant method to spatially map the transition dipole moment of emission from the first, second, and higher-order excited states, and directly compare them with each other to obtain a broader understanding of the underlying multiexcitonic interaction mechanisms.

Here, we report the direct measurement of the emission anisotropy of the BX state by coupling a defocused imaging setup to a two-dimensional (2D) SPAD array, and applying heralded post-selection of X and BX emission. This technique, dubbed Heralded Defocused Imaging, allows estimating the difference between the emission anisotropy of the X and BX emission transition dipole. The experimental results show higher anisotropy of the BX-to-X transition over the X-to-GS transition for type-I½ (quasi-type-II) CdSe/CdS and the opposite trend for type-II ZnSe/CdS seeded NRs. We rationalize these findings by discussing possible competing contributions from both the transient dynamics of the refractive index and the excitonic fine structure.

RESULTS

To investigate the transition dipole moment orientation of the BX-to-X transition we examine two nano-heterostructures that are known to emit partially polarized light from the X-to-GS transition. The first is CdSe/CdS seeded NRs and the second is ZnSe/CdS seeded NRs, termed here NR1 and NR2, respectively. Both heterostructures feature a quasi-spherical core (CdSe or ZnSe) within an elongated CdS rod shell. Figure 1a shows transmission electron microscope (TEM) images of the nanocrystals investigated in this work. The band alignment differences between the two nanocrystal types, illustrated in Figure 1b, result in qualitatively different charge-carrier wavefunctions, making them interesting candidates for this comparison. NR1 feature a charge carrier distribution characteristic of a type-I½ (also known as quasi-type-II) band alignment, where the hole is localized to the CdSe core and the electron is delocalized across the core and the rod. NR2 feature a charge-carrier distribution reflecting the type-II band alignment, where the hole and the electron are separated to the ZnSe core and CdS rod, respectively.[25] Notably, in both cases Coulomb attraction affects mostly the electron wavefunction distribution within the rod. The two seeded NR systems exhibit similar absorption profiles far from the band edge ($\lambda < 450$ nm, where absorption is dominated by the CdS rod) but different absorption profiles around the lowest excitonic peak, as shown in Figure 1c. The emission peaks of both systems are close to 600 nm. Synthesis and further details of the NCs can be found in section S1 of the supplementary information (SI).

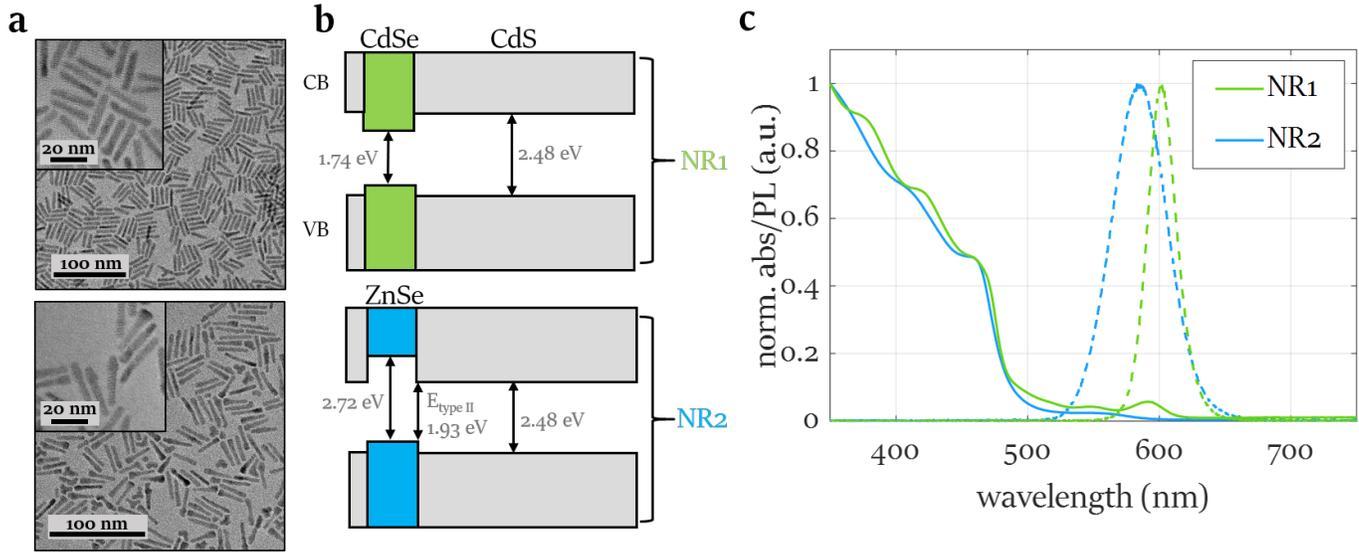

**Figure 1. Synthesis characterization.** (a) Transmission electron microscope (TEM) images for type-I½ CdSe/CdS seeded nanorods (NR1, top) and type-II ZnSe/CdS seeded nanorods (NR2, bottom). Insets are high resolution TEM images. Scale bars are 100 nm (20 nm in insets). (b) Energy diagrams of NR1 (top) and NR2 (bottom) showing the band gaps of all constituent materials and the type-II band-edge transition (bottom). (c) Normalized absorbance (abs, solid lines) and photoluminescence (PL, dashed lines) of NR1 (green) and NR2 (blue). Emission peaks are at ~600 nm and ~583 nm, respectively.

The experimental setup (figure 2a) is built around a commercial inverted microscope. A pulsed laser beam (70 ps, 470 nm, 5 MHz) is focused on a single particle by a high numerical aperture objective lens (x100, 1.3 NA). The linearly polarized laser is converted to circular polarization (using a $\lambda/4$ waveplate) to obtain uniform excitation of NRs, irrespective of orientation. Excitation intensity is well below saturation, at $\langle N \rangle \sim 0.1$ (mean number of photons absorbed per pulse, see SI section S2). Fluorescence emitted by the NR is collected *via* the same objective lens and filtered by a dichroic mirror and a long-pass filter. The tube lens is chosen such that the image is magnified by a factor of x75 and imaged onto a 23-pixel SPAD array detector (termed SPAD23), shifted by ~2 Rayleigh ranges from the image plane to create the defocused image. The shift magnitude and sign are crucial to create the defocused pattern as discussed in ref 1 and in SI section S3. The detector array contains 23 pixels (each pixel is an independent single-photon detector) arranged in a hexagonal lattice (see figure 2b, top) and connected to time-to-digital converters (TDCs) implemented on a field-programmable gate array (FPGA). Each detected photon is timestamped with a precision of ~100 ps (full width at half maximum) and address-stamped according to the pixel it was detected in. Intensity and temporal corrections are applied following refs 20,26 with adaptations to the SPAD23 detector and detailed in the SI section S4. Notably, the high temporal resolution allowed filtering detection pairs originating in inter-pixel optical crosstalk by temporal gating rather than *via* a statistical correction.

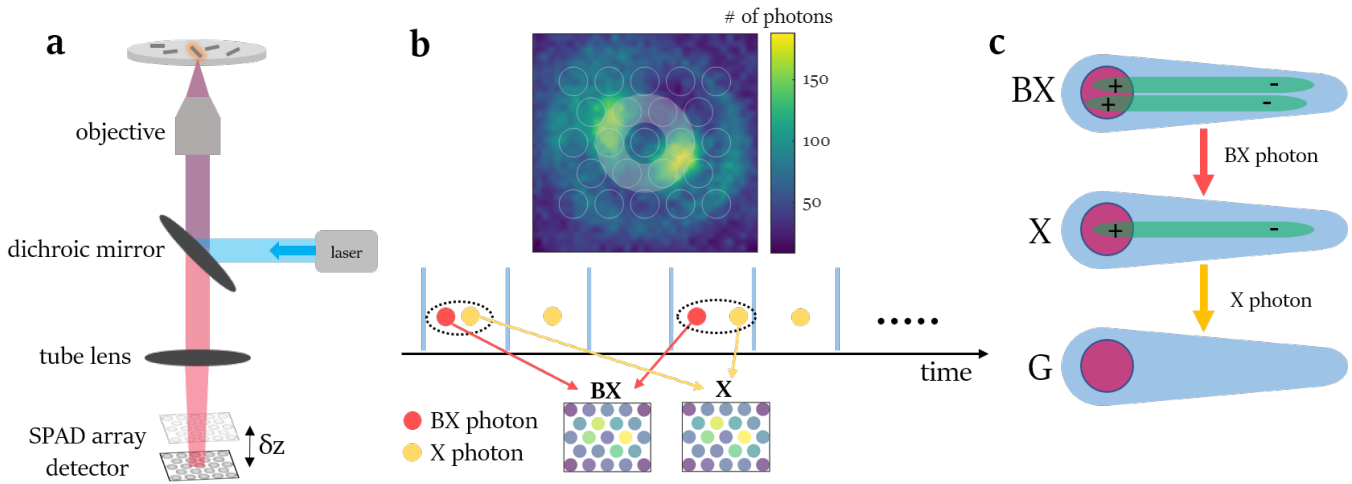

Figure 2. Optical setup and heralded defocused imaging technique. (a) Schematic of the defocused imaging setup consisting of an inverted microscope with a 470 nm laser excitation, a dichroic mirror, an objective lens, and a tube lens, coupled to a single-photon avalanche detector array of 23 pixels (SPAD23). Each detected photon is identified with a pixel number and an arrival time. (b) Top: defocused image of a single NR imaged with a CMOS camera. The overlaid 23 white circles represent the collection area of the 23 pixels of the SPAD array. The white shaded area highlights the six-pixel ring used for anisotropy estimation. Bottom: Illustration of the heralded spectroscopy analysis that post-selects photon pairs (indicated by black dotted ellipses) and sorts them into two groups: (i) first arriving photons (*i.e.* biexciton photons, filled red circles) and (ii) second arriving photons (*i.e.* exciton photons, filled orange circles). Summing each group of photons over all pixels yields emission maps of the biexciton and exciton states individually, as shown at the bottom. (c) A sketch of biexciton emission cascade of a doubly-excited seeded nanorod. The cascade features two subsequent transitions, defining the first emitted photon as a biexciton photon (red colored arrow) and the second emitted photon as an exciton photon (orange colored arrow).

The magnification (x75) was chosen such that the center ring of the defocused pattern falls on the six pixels around the detector center (highlighted in figure 2b). This ring contains the most information about the in-plane dipole orientation, and hence these six pixels are used in the subsequent analysis. Heralded isolation of BX and X emission was done following the scheme described in refs 20,21. Briefly, the high temporal resolution of the detectors allows isolating emission cascades originating in cascaded relaxation from the BX to the X to the GS state (BX-X-GS, illustrated in figure 2c) from the overwhelmingly stronger singly-excited fluorescence background, by post-selecting photon pairs detected following the same excitation pulse (figure 2b, bottom panel). The first photon of the pair is associated with the BX-to-X transition, and the second with the X-to-GS transition. Additionally, pairs with inter-detection delay shorter than 4 ns or where the first photon arrived more than 2 ns after the excitation pulse are filtered, to minimize the number of artificial pairs induced by optical crosstalk and dark counts, respectively (see SI section S4).

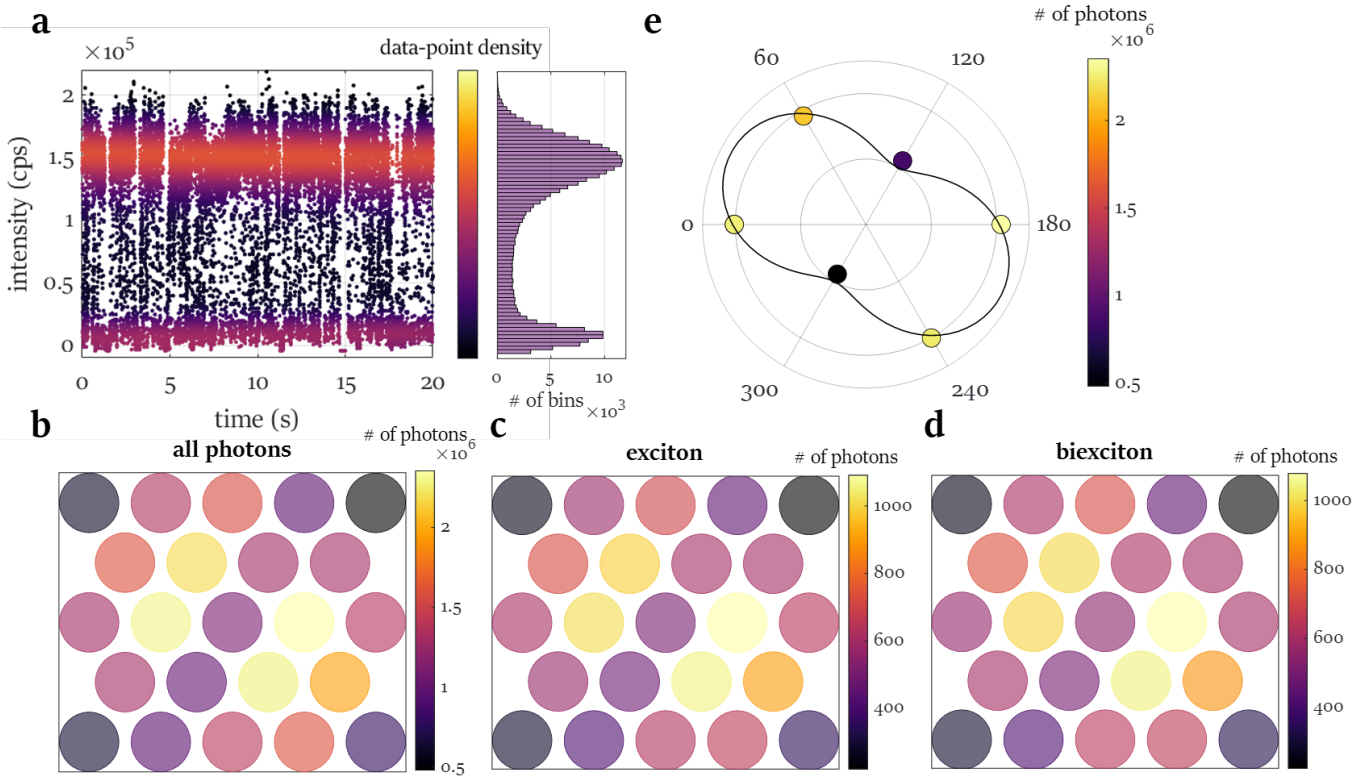

**Figure 3. Heralded defocused imaging analysis of a single nanorod.** (a) Left: total fluorescence intensity collected for all pixels as a function of measurement time in a 20 second time window. Each data-point represents the average intensity over a 1 ms time

bin, colored according to the local density of data-points for clarity, and corrected for detector dark counts. Right: histogram of intensity values over a 5 min measurement revealing the "on" and "off" states, evident as peak occurrences at high and low intensity, respectively. (b-d) Histograms, by pixel, of all detected photons (b), and post-selected exciton (c) and biexciton (d) detections from a 5 min measurement of a single type-II ZnSe/CdS seeded nanorod (NR2), applying heralded defocused imaging. Color scale represents the number of detections at a given detector pixel. (e) Polar representation of the intensity values detected by the six pixels of the inner ring of the array (highlighted in figure 2b, top) for all detected photons in the measurement, along with the fit (black solid line) to the integrated dipole emission model. The six data-points are colored according to the number of detected photons in each pixel. The same fitting process is applied for the exciton and biexciton data sets and presented in section S5 of the SI.

Figure 3 presents the results of the analysis described above for a 5-min measurement of a single representative NR. Figure 3a depicts the fluorescence intensity as a function of time for 20 seconds of the measurement, summed over all detector pixels. The contribution of dark counts is statistically estimated from prior characterization of the detector and subtracted from the raw measured intensity. The intensity trace shows a typical single-particle behavior, with fluctuations between well-defined "on" and "off" states. Figure 3b visualizes the intensity measured in each of the 23 pixels of the array, integrated over the whole measurement. Notably, the anisotropy of the defocused pattern seen in figure 2b is also evident in figure 3b, despite the lower spatial sampling resolution. To quantify this anisotropy, and hence the level of polarization, the anisotropy of the six-pixel ring around the central pixel is estimated by a fit to a sum of a squared sine and an isotropic background:

$$I(\theta) = A \cdot \left( \frac{a}{\pi} \cdot \sin^2(\theta - \phi) + \frac{1-a}{2\pi} \right) \qquad (1)$$

$I(\theta)$ is the photon count rate (intensity) as a function of $\theta$, the angle along the six-pixel circle, $A$ is a normalization factor (the total number of photon pairs detected), $a$ is the anisotropy, and $\phi$ is a global phase. The fit parameters are $a$ and $\phi$. The anisotropy value, $a$, is used as a measure of the emission anisotropy. While $a$ is not a direct estimator of emission polarization, it is correlated with emission polarization and thus allows us to compare the dipole orientation and magnitude of the BX and X states as described below. To take into account the fact that the collection area of each pixel subtends roughly one sixth of the circle, the values measured by the array pixels are fitted to the integral of eq.1 over ±30°, which describes the photon count rate of a pixel detector centered at $\theta'$:

$$I(\theta') = A \cdot \left(\frac{1}{6} - a \cdot \frac{\sqrt{3}}{4\pi} \cdot \cos(2 \cdot (\theta' - \phi))\right) \qquad (2)$$

The fit results for the specific NR featured in figure 3e, considering all detected single photons, are: $a = 0.48 \pm 0.04$ and $\phi = 118° \pm 3°$ (all errors in the paper are estimated as the 68% confidence interval of the fit). The angle $\phi$ extracted from the fit features the in-plane orientation of the transition dipole, which is aligned along the NR, and conforms with the emission patterns appearing in figures 3b-d.

Applying the heralded isolation of BX-X-GS emission cascades as described above (figure 2b, bottom) to the same measurement yields ~11,000 post-selected BX-X pairs. The first photon of each pair can be associated with BX emission, and the second with emission from the X state. Repeating the emission pattern anisotropy estimation for the post-selected X-to-GS and BX-to-X yields the histograms shown in figures 3c and 3d, respectively. The fit results for the BX and X are: $a_{BX} = 0.44 \pm 0.05$, $\phi_{BX} = 119° \pm 3°$, $a_X = 0.45 \pm 0.05$, and $\phi_X = 118° \pm 3°$, respectively. Two immediate observations are that the anisotropy and transition dipole orientation are similar, to within the error of the fit. However, as shown below, the slightly higher anisotropy of the X emission is characteristic for this type of NR and becomes evident upon ensemble averaging. Note that since the X emission dominates the overall emission, we expect to get similar anisotropy values for the X ($a_X$) and for all the photons in the measurement ($a$). The pixel dead-time leads to an apparent reduction in both the X and BX anisotropy values due to pairs of photons that were supposed to be detected on the same pixel. The reported X and BX anisotropy values in figures 3 and 4 are corrected for this (see SI section S4) and correlate better with the anisotropy values extracted for all the photons detected in a measurement. Importantly, this correction affects the X and BX anisotropy values equally and does not affect the difference between them. It is worth noting that the high-temporal resolution single-photon detection allows for many more post-analyses and insight from the same raw data. Other examples of single-particle analyses of NR1 and NR2 can be found in SI section S6.

The heralded single-particle analysis described above was repeated for 28 NR1 and 79 NR2 nanocrystals. The aggregate results are presented in figure 4a. It is evident that the anisotropy of BX and X emission ($a_{BX}$ and $a_X$) is highly correlated. All single-NR measurements are within the fit error from the diagonal, representing equal anisotropy of the BX and X emission, indicating similar anisotropy of the respective transition dipole moments. The orientation of the BX-to-X and X-to-GS transition dipoles ($\phi_{BX}$ and $\phi_X$), are also aligned, within the fit error, for all single NRs (see SI section S7). This similarity can be expected, given the twofold degeneracy of the lowest excited state in these NCs.[27] However, exciton—exciton interaction can somewhat alter the properties of the BX emission, as is well known for the BX emission spectrum.[20] Indeed, further analysis of the results, clustered by NR type, reveals small, but statistically significant deviations of the BX-to-X transition dipole anisotropy from that of the X-to-GS transition. Aggregate analysis of NR1 (figure 4b) reveals that BX emission is typically more anisotropic than the X for NRs of this type. The statistical significance of this observation is confirmed *via* a paired Student's t-test, yielding a score of 2.5 (corresponding to a p-value of ~0.02). NR2 show the opposite trend, where BX emission is less anisotropic than X emission (figure 4c), with a paired Student's t-test score of 3.8 (corresponding to a p-value of ~0.0003).

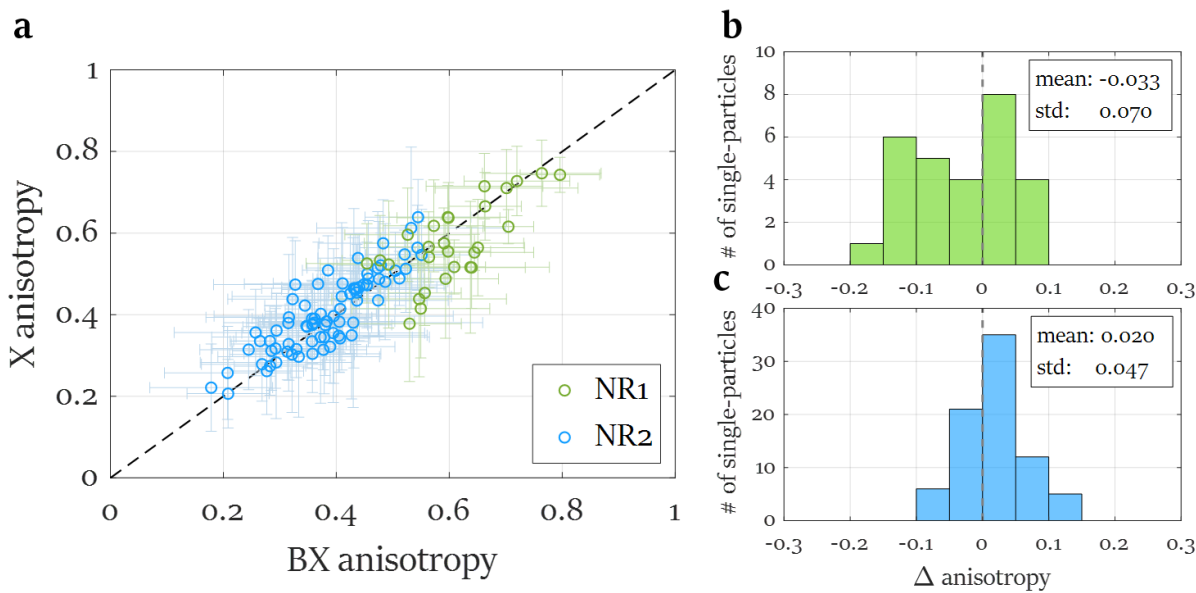

**Figure 4. Aggregate X and BX emission anisotropy analysis.** (a) Exciton emission anisotropy versus biexciton emission anisotropy, extracted by heralded defocused imaging, for all measured single NCs: 28 type-I½ CdSe/CdS (NR1, green rings) and 79 type-II ZnSe/CdS (NR2, blue rings) seeded nanorods. The dashed black line in all panels is a guide to the eye, indicating the same anisotropy values for both X and BX photons. (b-c) Histograms of $\mathbf{\Delta}_{anisotropy} = \mathbf{X}_{anisotropy} - \mathbf{BX}_{anisotropy}$ for NR1 (b), and NR2 (c).

DISCUSSION

The experimental results presented above indicate a small yet statistically significant deviation of the anisotropy of the BX-to-X emission from that of the X-to-GS emission. In the following we discuss two mechanisms governing X-to-GS emission anisotropy: dielectric anisotropy and exciton fine structure. We propose how these mechanisms might lead to a deviation of the BX-to-X emission anisotropy from that of the X-to-GS one, leading to the observed results.

The dielectric contribution to emission anisotropy in elongated NCs was discussed extensively in past literature, relating the increased probability to emit light polarized parallel to the long axis of the NR to the different boundary conditions of the electric field parallel and perpendicular to the rod surface.[6,28–30] This effect can be estimated by approximating the NRs as homogeneous dielectric ellipsoids using an effective medium calculation.[31–33] However, the emission anisotropy of the BX state is different as light emission occurs against an existing spectator exciton background. The absorption bleach induced by the presence of the X implies a change in the effective refractive index of the NR. Indeed, Tanghe and co-workers recently reported on a sizable phase modulation of light passing a sample of CdSe nanoplatelets (NPLs) upon excitation of the nanoplatelets, which is proportional to a refractive index change. In their work, TA measurements were translated into transient refractive index data, estimating the refractive index modulation shortly after excitation. Upon photon absorption, the imaginary part of the complex refractive index, the extinction coefficient, is transiently changed. The Kramers-Kronig relations, connecting the real and imaginary parts of analytic complex functions, imply that also the real part of the refractive index must therefore change.[34] As a result, the dielectric response of the material is altered according to the relation $\tilde{n} = \sqrt{\tilde{\varepsilon}}$ ($\tilde{n}$ and $\tilde{\varepsilon}$ are the complex refractive index and complex dielectric constant, respectively). Tanghe *et al.* considered this effect in several spectral bands with respect to the bandgap of the nanoplatelets, and showed that photoexcitation of their material increases the refractive index on the high energy side of the absorption bleach (*i.e.* above the bandgap) and decreases it on the low energy side. Here, despite the difference in the NC structure and composition, their results may be used as a guideline to estimate the magnitude of this effect. Although the results of Tanghe *et al.* were obtained for nanoplatelets, we can use them to provide a rough estimate of the magnitude of expected change. In their reported results, the refractive index change ($\Delta n$) at the level of 1 excitation per NPL ($\langle N \rangle = 1$) measured at 600 nm (almost 100 nm red-shifted from the absorption edge) is ~0.01. In our particles, both NR1 and NR2, the emission wavelength of 600 nm is also redshifted by ~100 nm from the main absorption edge, hence we expect a similar change of refractive index due to photoexcitation. Considering the dielectric parameters of CdS, the relative change in the refractive index corresponds to $\frac{|\Delta n|}{n_0} \approx 0.4\%$ for both NR1 and NR2 (since both have a CdS rod shell and the refractive index, $n_0$, of CdS at 600 nm is 2.34[6]). The dielectric contribution to the anisotropy scales with square of the ratio of the internal electric field strength between the major and minor axis of the rod ($R_e = \frac{E_\parallel}{E_\perp}$). Following the derivation of Vezzoli *et al.*[6] (see also section S8 in the SI), the relative change in the dielectric effect on emission anisotropy, $\frac{|\Delta R_e^2|}{R_e^2}$, can be estimated to be ~1%. The BX emission anisotropy is expected to be reduced by this amount because the spectator X decreases the CdS refractive index at 600 nm.

In addition to the dielectric effect, the exciton fine structure also influences the emission polarization. The intrinsic difference between the crystal structure of CdSe and ZnSe may explain the qualitatively different results for the two NR types. As both CdSe and CdS have a wurtzite crystal structure with hexagonal symmetry, the c-axis of the CdSe core aligns with that of the CdS shell.[6,35,36] This may induce a stronger linear polarization of the emission along the rod, as evident in the higher anisotropy values for NR1 (figure 4a). The exciton fine structure of CdSe was extensively studied in previous works.[6,25,31,37–41] The eightfold degenerate exciton ground state of a spherical NC is greatly influenced by the morphology, crystal structure anisotropy, and electron-hole exchange interaction. All mentioned factors may alter the splitting, ordering, and transition oscillator strengths of the states.[38] The shape anisotropy, together with the exchange interaction, split the eightfold degenerate exciton into five levels, three of them are twofold degenerate. For elongated CdSe NRs, exchange interactions make $0^u$ the lowest optically active exciton state.[6] The $0^u$ state features the observed 1D dipole emission, aligned with the elongated dimension of the NR. It can be assumed that for NR1, the overlap between the hole and electron wavefunctions of the BX is stronger than that of the X due to the stronger Coulomb attraction to the two holes confined in the CdSe core. The higher overlap may increase the exchange interaction, and hence the splitting of the fine structure. Enhanced splitting means emission will be even more dominated by the lowest fine structure state, thus increasing BX anisotropy. The effect of the fine structure on the BX emission anisotropy is much smaller in the case of NR2 due to the charge separation of the excited electron and hole, which dramatically reduces the electron-hole exchange interaction. To summarize this comparison, the dielectric effect is expected to reduce the anisotropy of the BX (since emission is red-shifted from the absorption edge), whereas the increased electron-hole overlap of the BX is expected to increase the anisotropy. The former has a similar influence on both type-I½ and type-II NRs, while the latter predominantly affects the type-I½. This might explain why the type-I½ NR1 shows increased BX emission anisotropy compared to the X (fine-structure dominated), while the type-II NR2 has decreased BX anisotropy (dielectric-effect dominated). Interestingly, we have also studied a sample of shorter type-II NRs (see SI section S9), which exhibited a similar trend as the longer type-II NRs (NR2) showing a slightly lower emission anisotropy for the BX compared with the X. Due to lack of statistics, no significant conclusion can be deduced for these NRs.

It is worth noting that other methods can also probe the photon polarization correlation in emission cascades. A seminal example is a work by Aspect *et al.*, demonstrating a violation of Bell's inequalities in the polarization correlation of photon pairs emitted in a radiative cascade of calcium.[42] The system applied there consists of a beamsplitter splitting light into two paths; each path is further divided between two single-photon detectors by rotatable polarizing beamsplitters, allowing estimating and correlating the polarization in the two arms. While extracting quantitative polarization values is more challenging with heralded defocused imaging as performed here, it introduces several significant advantages over the abovementioned technique. Heralded defocused imaging obviates the need to repeat the experiment at multiple polarizer orientations as it is sensitive to all polarization orientations at once. The detection and timing setup, consisting of a single compact component (the SPAD array detector) placed at the defocused image plane, is significantly simpler than traditional multiplexed polarimeters. Finally, the most critical advantage of heralded defocused imaging is its potential for straightforward adaptation to higher-pixel count SPAD arrays. This up-scaling will allow extracting the dimensionality and 3D orientation of multiexcitonic transition dipoles, as previously demonstrated with imaging detectors for the singly-excited state.[1,16,43]

CONCLUSIONS

We present a new approach to directly probe the transition dipole moment of the BX-to-X transition in single NCs. Heralded defocused imaging provides us with the dipole orientation mapping of single NRs onto a two-dimensional SPAD array detector to temporally differentiate between the emission transition dipole moment of the first and second excited states. The results reveal variations between the emission transition dipole moment of the two first excitonic states, showing higher BX emission anisotropy for the type-I½ NRs and lower BX emission anisotropy for the type-II NRs compared with the X emission anisotropy. We discuss possible transient deviations of refractive index and exciton fine structure present in the BX state, and how they may explain these observations. The heralded defocused imaging technique introduced here expands the limited set of experimental tools that directly probe multiply-excited states in single semiconductor nanocrystals. It can be easily realized, applied to various nanocrystal systems, and scaling up the number of detector pixels can further support three-dimensional polarization analysis. Heralded defocused imaging demonstrates the benefit of harnessing photon correlations to investigate multiply-excited states in semiconductor nanocrystals by uncovering previously inaccessible insights, and holds a great potential to further our understanding of these materials.

ASSOCIATED CONTENT

**Supporting Information**.

Details of the nanorods including synthesis, characterization methods, sample preparation and saturation estimation assay; details of the defocused single-particle spectroscopy setup; dark count rate correction, crosstalk estimation, and detector dead-time correction details; complementary anisotropy analysis for the exciton and biexciton; additional examples of single-particle measurements; exciton-biexciton in-plane angle correlation details; dielectric effect estimation details; details of additional measured sample

AUTHOR INFORMATION

Corresponding Author

*Dan.Oron@weizmann.ac.il

Author Contributions

†These authors contributed equally.

Funding Sources

This research was supported by the Israel Science Foundation (ISF), and the Directorate for Defense Research and Development (DDR&D), grant No. 3415/21. D.A. gratefully acknowledges support by the VATAT Fellowship for female PhD students in Physics/Math and Computer Science. D.O. is the incumbent of the Harry Weinrebe professorial chair of laser physics.

Notes

The authors declare no competing financial interest.

ACKNOWLEDGMENT

The authors would like to thank Ivan Michel Antolovic and Harald Homulle from "Pi Imaging Technology" for a fruitful collaboration and technical support in implementing the SPAD23 system, and Ariel Halfon and Nadav Frenkel for assisting with supporting measurements and characterization of the SPAD23 detector.

# Supporting Information: Resolving the emission transition dipole moments of single doubly-excited seeded nanorods *via* heralded defocused imaging

*Daniel Amgar[1,†], Gur Lubin[1,†], Gaoling Yang[2], Freddy T. Rabouw[3], and Dan Oron[4,*]*

[1] Department of Physics of Complex Systems, Weizmann Institute of Science, Rehovot 76100, Israel

[2] School of Optics and Photonics, Beijing Institute of Technology, China

[3] Debye Institute for Nanomaterials Science, Utrecht University, Princetonplein 1, 3584 CC Utrecht, The Netherlands

[4] Department of Molecular Chemistry and Materials Science, Weizmann Institute of Science, Rehovot 76100, Israel

## S1: Materials and methods

### Details of the dot-in-rods synthesis

ZnSe/CdS dot-in-rod nanocrystals were synthesized following a previously published procedure with some modifications.[1]

### Chemicals

Hexadecylamine (HDA, 98%, Aldrich), diethylzinc (Et$_2$Zn, 1 M solution in hexane, Aldrich) Cadmium oxide (99.99%, Aldrich), Sulfur (99.999%, Aldrich), Selenium (99.999%, Aldrich), oleic acid (OA, 90%, Aldrich), trioctylphosphine (TOP, 90%, Aldrich), dodecylamine (98%, Fluka), trioctylphosphine oxide (TOPO, technical grade, 99% Aldrich), 1-octadecene (ODE, technical grade, 90% Aldrich), hexylphosphonic acid (HPA, 99%, PCI), n-octadecylphosphonic acid (ODPA, 99%, PCI), methanol (anhydrous, 99.8%, Aldrich), hexane (anhydrous, 99.9%, Aldrich), toluene (99.8%, Aldrich). All chemicals were used as received without any further purification.

### Synthesis of ZnSe NCs

9.4 g of HDA was degassed under vacuum at 120 °C in a reaction flask, under argon flow, the mixture was heated up to 310 °C. Then, a mixture of 1 mL 1.0 M selenium dissolved in TOP, 0.8 mL diethylzinc and 4 mL TOP, was quickly injected. The reaction was continued at a constant temperature of 270 °C for 25 min and then cooled to room temperature.

### Preparation of Cadmium and Sulfur Stock Solutions

0.034 M cadmium oleate was prepared by mixing 0.03 g (0.24 mmol) CdO in 0.6 mL oleic acid and 6.4 mL ODE. The solution was heated to 280 °C under argon flow with rigorous stirring until all of the CdO dissolved. 0.29 M S solution was prepared by adding 23.3 mg sulfur in 2.5 mL of dodecylamine at ~40 °C.

### CdS Shell Synthesis

For typical CdS shell coating, 1.1 g unprocessed ZnSe cores, 5.3 mL octadecene (ODE) were loaded into a 50 mL three-neck flask. The solution was degassed at 100 °C. After that the solution was heated to 240 °C under argon, a mixture of 0.6 mL of 0.034 mmol/mL cadmium oleate stock solution and 0.06 mL of 0.29 mmol/mL sulfur stock solution was injected continuously at 0.72 mL/h. After the injection was finished, the mixture was further annealed for 5 min at 240 °C and then cooled down to room temperature.

ZnSe/CdS-CdS NRs Synthesis

This synthesis was adapted from the previously reported procedure in the literature.[2,3] In a typical synthesis CdO (60 mg), ODPA (290 mg) and HPA (80 mg) are mixed in TOPO (3.0 g). The mixture is degassed under vacuum at 150 °C for 90 min. After degassing step, the solution was heated to 380 °C under argon until it became clear, then 1.8 mL of TOP was injected and the temperature was recovered to 380 °C. Subsequently a solution of 120 mg S in 1.8 mL TOP mix with 40 nmol ZnSe/CdS nanocrystals is rapidly injected. Then the growth was stopped after nanorods grow for 8 min at 365 °C. The NRs were precipitated with methanol and dispersed in toluene.

Synthesis of CdSe NCs

TOPO (3.0g), ODPA (0.280g) and CdO (0.060g) are mixed in a 50mL flask, heated to 150°C and exposed to vacuum for 1.5 hour. Then, under nitrogen, the solution is heated to above 370°C to dissolve the CdO until it turns optically clear and colorless. At this point, 1.8 mL of TOP is injected in the flask and the temperature is allowed to recover back to 370°C. At this temperature, Se:TOP solution (0.058g Se + 0.5 mL TOP) was injected and the heating mantle is removed to stop the reaction after 2 minutes. After the synthesis, the nanocrystals are precipitated with methanol, they are washed by repeated redissolution in toluene and precipitation with the addition of methanol, and they are finally dissolved in toluene.

## Synthesis of CdSe/CdS dot-in-rod

This synthesis was adapted from the previously reported procedure in the literature.[3] Firstly, 0.09 g CdO, 3 g TOPO, 0.08 g HPA and 0.290 g ODPA were dissolved and vacuumed in advance at 150°C. The resulting solution was allowed to be heated to 350 °C under nitrogen and 1.5 g TOP was injected. Then the mixture was heated to 380 °C and propriate amount of CdSe seed together with 0.12 g S in 1.8 mL TOP were injected simultaneously. The length of CdS was controlled by adjusting the concentration of CdSe seed. The mixed solution was maintained at ~380 °C for 8 min under nitrogen to complete the growth of the CdSe/CdS dot-in-rods, and the heating mantle was moved to stop the reaction. The CdSe/CdS dot-in-rods were further purified with toluene and methanol as solvent and nonsolvent, then dispersed and stored in toluene.

## Characterization methods

TEM images were taken on a JEOL 2100 TEM equipped with a LaB6 filament at an acceleration voltage of 200 kV on a Gatan US1000 CCD camera. UV-vis absorption spectra were measured using a UV-vis-NIR spectrometer (V- 670, JASCO). The fluorescence spectrum was measured using USB4000 Ocean Optics spectrometer excited by a fiber coupled 407 nm LED in an orthogonal collection setup.

## Sample preparation

The samples for the single-particle experiments were prepared as follows. A stock solution of DiR nanocrystals was diluted in a solution of poly(methyl methacrylate) (PMMA, Aldrich) in Toluene by $10^4 - 10^6$. The PMMA/Toluene solution was prepared by dissolving 3 wt% of PMMA powder in Toluene. The solution was gently stirred and heated to ~65° overnight. To prepare the sample for the microscope, 200 µl of the PMMA/Toluene solution were spin coated onto a glass coverslip in two steps; (i) 5 seconds at 800 rpm, (ii) 45 seconds at 2000 rpm.

## S2: Saturation experiments

To avoid possible contamination of the results with emission from higher-multiexcitonic states (triexcitons and above), illumination power was kept far below saturation. At these illumination intensities the probability of absorbing 3 photons or more in a single pulse and a radiative decay of the triexciton or higher states is negligible. The saturation power, for which $\langle N \rangle$, the average number of photons absorbed per pulse, equals unity, was estimated following the scheme described in the supporting information of ref 4: A single particle was illuminated by the same setup described in the main text. During the measurement, the laser power was increased every 10 s, in 10 equally-spaced power steps up to some maximal power and then lowered back down to the starting point. In the measurement featured in figure S1, the power range was from $\sim 20 \ nW$ to $\sim 700 \ nW$. Figure S1a presents the intensity trace for a representative measurement showing the ascending and descending intensity as a function of measurement time with the power steps, as well as the 'on'-'off' blinking mentioned in the main text (in relation to figure 3a). To construct the saturation curve, we considered both time windows with the same power (from the ascent and descent) and extracted the 'on' state typical intensity for that illumination power (the peak occurrence intensity, seen as the brighter, higher-density regions in figure S1a). This isolation of the 'on' state peak is aimed to avoid bias associated with possibly enhanced blinking at higher intensity excitations, as described in the supporting information of ref 4 The saturation curve is shown in figure S1b with a fit (solid red line) to a saturation function:

$$P = a \cdot \left(1 - e^{-\frac{I}{I_{sat}}}\right), \tag{S1}$$

Where $P$ is the 'on' state peak, $a$ is asymptotic 'on' state peak, $I$ is excitation power, and $I_{sat}$ is the saturation power. The data in figure S1b agree well with the suggested model and the saturation power extracted from the fit is $I_{sat} = 670 \pm 70\ nW$ (68% confidence interval, green dashed line). The excitation power used in this work is $I_{exc} = \sim 80\ nW$ (purple dashed line in figure S1b), which is far below the saturation power.

The probability to create $n$ excitons per excitation pulse can be estimated from the Poissonian distribution:

$$(n) = \frac{\lambda^n \cdot e^{-\lambda}}{n!} \tag{S2}$$

$$\lambda = \frac{I}{I_{sat}} \tag{S3}$$

The probability that the NC will absorb at least one photon in an excitation pulse, at the excitation power used in this work, is thus ~11%. The probability to absorb at least two photons per pulse (creating a biexciton) is 0.66%, and of absorbing three or more photons is 0.026%.

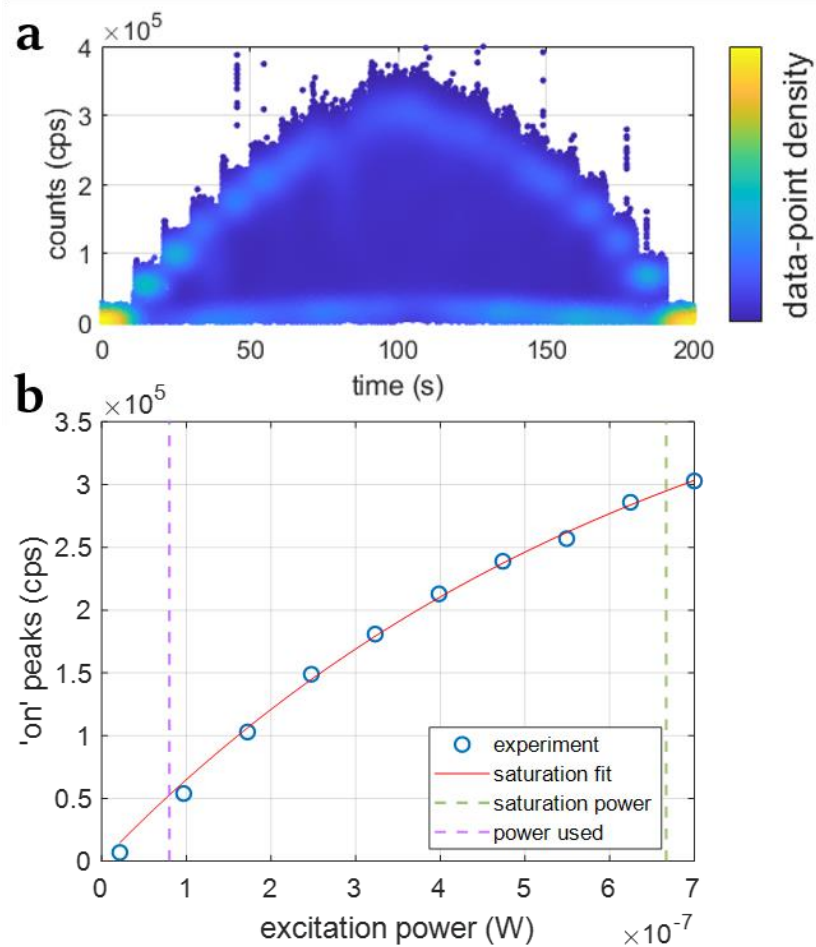

**Figure S1. Saturation assay.** (a) Intensity trace as a function of measurement time with 10 equally-spaced power steps ranging from 20 nW to 700 nW, where each step lasts 10 seconds. The power is gradually increased up to the maximal power and decreased back to the starting power. (b) Saturation curve presenting the "on" peak intensity as a function of excitation power fitted to a saturation function (red solid line). The saturation power and the power used in the experiments are marked by green and purple dashed lines, respectively.

## S3: Defocused single-particle spectroscopy setup

The experimental setup (figure 2a in the main text) is built around a commercial inverted microscope (Eclipse Ti-U, Nikon). A pulsed diode laser (LDH-P-C-470B, PicoQuant) provides a collimated beam at a wavelength of 470 nm and a repetition rate of 5 MHz. The beam is focused by a high numerical aperture (NA) oil immersion objective lens (x100, 1.3 NA, Nikon), which also collects the resulting fluorescence light. An additional magnification step inside the microscope results in magnification of x150. Back-scattered laser light is filtered out by dichroic mirror (FF484-FDi02-t3, Semrock) and a long pass dielectric filter (BLP01-473R, Semrock). The defocused imaging path contains a 75 mm focal length relay lens that de-magnifies the image by a factor of two, for an overall magnification of ×75. An equivalent imaging path leading to a complementary metal-oxide-semiconductor (CMOS) camera (BFS, FLIR) is available using a flip mirror, allowing wide field imaging used for imaging the sample and aligning a single NC with the excitation beam before each measurement. Finally, the desired defocused emission pattern is imaged onto a 23-pixel single-photon avalanche diode (SPAD) array, fabricated in CMOS technology (SPAD23, Pi Imaging Technology), with an estimated diffraction-limited spot of ~20 μm and a defocused spot diameter of ~ 220 μm. The SPAD is attached to an XYZ stage to modify the defocus distance and for alignment. A field-programmable gate array (FPGA) with a coarse clock system, and an implemented array of time-to-digital converters (TDCs) with a fine resolution of ~10 ps (both synchronized with the laser excitation), assigns timestamps and pixel addresses to single-photon detections in the 23 pixels of the array. The trace of detections is then analyzed by a dedicated MATLAB script, implementing temporal and intensity corrections and analysis schemes.

## S4: Corrections

As described previously in refs 4–6, two sources of artificial photon detection pairs must be considered when analyzing photon correlations with SPAD arrays. The first is pairs where at least one detection originates from the detector's dark counts rather than a fluorescence photon. The second is where the false photon pair detections arise from inter-pixel crosstalk. As described below, both sources of artificial pairs feature very different temporal responses from 'true' BX-X photon pairs and hence can be filtered by applying temporal gating. Residual dark count-induced artificial pairs are corrected statistically. Additionally, 'true' BX-X photon pairs that impinged on the same detector pixel are not registered by the system due to pixel dead time. The number of these undetected pairs is estimated from the measurements and added to the results. These three intensity corrections are discussed below.

### Dark counts

The SPAD array detector features some probability for false photon detections, even when no fluorescence photons impinge on the sensor. These detections are known as dark counts. In this work, the dark count rate (DCR) is negligible compared to the single-photon detection rate (except for one 'hot' pixel) but comparable to the estimated rate of detected BX-X photon pairs.

Unlike the signal probability that decays exponentially with the delay from each excitation pulse, DCR is time-independent. Hence, the time-gating described in the main text filters out a significant portion of dark counts, while retaining almost all of the fluorescence data, leading to a better signal-to-noise ratio (SNR). Specifically, only photons arriving within the first ~2 ns following any excitation pulse are considered for labelling as BX emission (due to the fast decay of the doubly-excited state). As the laser period is 200 ns, and dark counts are time-independent, this filters out ~99% of the potential dark count-induced detection pairs, where the BX is a dark count. Similarly, considering only detections within

~100 ns following a BX detection to label as X emission filters out ~50% of the dark count-induced detection pairs, where the X is a dark count.

The residual dark count-induced pairs left after temporal gating are corrected by the following scheme. Each pixel's DCR is pre-characterized by performing a 100 s intensity measurement in the dark (figure S2). For each single-NR measurement presented in the main text, the temporally-resolved intensity in each pixel is extracted by a standard time-correlated single-photon counting (TCSPC) analysis. Next, the number of false photon pairs consisting of at least one dark count is estimated for each measurement. This estimation is done by multiplying the probabilities of 'true' photon detections and dark count events to occur, for each pixel pair, within the temporal gates employed for the heralded post-selection described above. For the measurement featured in figure 3 of the main text, we estimate that out of the total 14535 detected pairs, the BX detection was a dark count in ~658, the X detection was a dark count in ~239, and both X and BX were dark counts in ~2. Hence, the temporal gates helped mitigate the dark-count-induced pairs to under an order of magnitude below the BX-X pairs signal. Results shown in the main text are after subtracting these estimated values of dark-count-induced detection pairs from the raw measurement data of each pixel pair. The significance of minimizing the impact of dark counts (by temporal gating) before this statistical correction, is to avoid the additional shot noise associated with the higher photon counts. Since shot noise is isotropic, it can skew our estimation of anisotropy values for the X and BX.

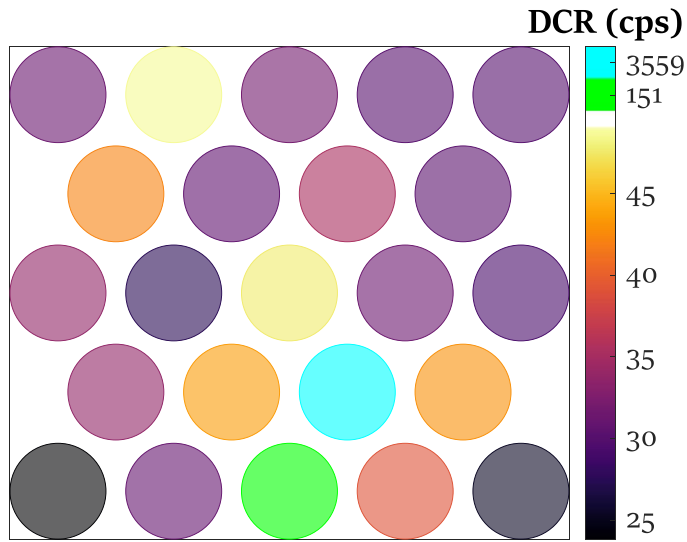

**Figure S2. Dark count rate (DCR).** The dark count rate of the detector pixels in counts per second (cps). The median DCR is ~32 cps. Note the 'hot' pixel at the second-to-bottom row with 2 orders of magnitude higher DCR than the median.

## Crosstalk

The densely packed pixels give rise to inter-pixel optical crosstalk, attributed to photons emitted due to the detection process in one pixel and then detected by a neighboring pixel.[7] The SPAD array design mitigates crosstalk significantly[8], but the small residual crosstalk probabilities are typically overwhelmingly higher than the probabilities for 'true' BX-X detections.[5] However, crosstalk and BX-X pairs feature separate time scales of inter-detection delay. Typical delays between a 'true' detection and the resulting crosstalk detection correspond roughly to the temporal precision of the detector (~100 ps FWHM), which is considerably shorter than the typical delay between BX and X emission, corresponding to the X fluorescence lifetime ($\tau > 10$ ns). Hence the requirement for at least 4 ns inter-detection delay applied in the heralded post-selection (see main text) filters almost all crosstalk detection pairs and just a small fraction of the BX-X pair signal.

The number of crosstalk-induced pairs in each measurement was estimated following the procedure detailed in ref 5, with some modifications. Briefly, to characterize crosstalk probabilities, the detector array is illuminated by a thermal light source (a halogen lamp), and the second-order correlation of photon arrival times, $G^{(2)}$, is extracted. The $G^{(2)}$ curve is expected to be flat for such a classical light source; however, a sharp peak is evident at short time-delays attributed to inter-pixel crosstalk. The excess photon pairs detected at short time delays are extracted for each pixel pair. These values are then divided by the number of overall single-photon detections in the pixel of the first detection to derive the time-resolved crosstalk probability for each pixel pair. Time-resolved crosstalk probability is the probability that, given a detection in one pixel, another pixel will register a crosstalk event at a given time delay. These probabilities are shown in figure S3a, where the rapid decay with time is evident. Figure S3b illustrates the effect of temporal gating in mitigating crosstalk contribution by plotting the crosstalk probability as a function of the inter-detection temporal gate (the minimal BX-X delay considered for the heralded analysis). For the gate used in this work (4 ns, dashed line in figure S3b), the overall crosstalk probability is $\sim 2.28 \cdot 10^{-4}$ per detected photon, almost two orders of magnitude less than without gating. For the measurement shown in figure 3 of the main text, this translates to ~380 crosstalk pairs mistakenly labelled as BX-X cascades by the heralded post-selection ($\sim 1.66 \cdot 10^6$ single photons were detected within the 2 ns following each excitation pulse; $1.66 \cdot 10^6 \cdot 2.28 \cdot 10^{-4} \approx 380$). We consider this value negligible compared to the overall 14535 post-selected BX-X cascades.

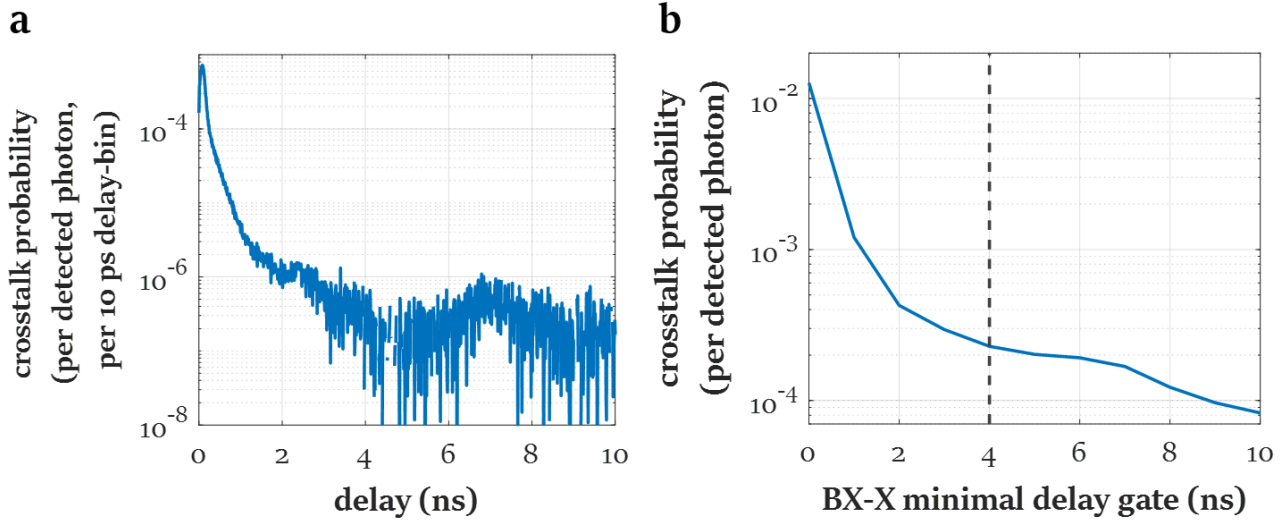

**Figure S3. Crosstalk characterization.** (a) Temporally resolved crosstalk probability. The vertical axis represents the probability for a crosstalk detection following any detection by the array, at the delay indicated on the horizontal axis. Delay-bins are 10 ps. (b) The crosstalk probability after applying the temporal gating as part of the heralded post selection. The vertical axis represents the probability for a crosstalk event to occur following a detection, at a delay longer than that indicated by the horizontal axis. This is generated by integrating the values in panel (a) from the BX-X minimal delay gate to infinity. Dashed line represents 4 ns, the temporal-gate used in this work, corresponding to a crosstalk probability of $\sim 2.28 \cdot 10^{-4}$, about two orders of magnitude below the crosstalk probability without gating ($\sim 1.27 \cdot 10^{-2}$).

Note that the critical difference from the crosstalk analysis in ref 5 is the detector's higher temporal precision, which allows resolving the order of detections in the crosstalk characterization and their inter-detection delay. This resolution is what allows the temporal gating described above instead of a statistical correction. It also alleviates the need to assume symmetric crosstalk probabilities, as done in ref 5.

## Same pixel pairs

If the BX and the X photons impinge on the same detector pixel, the second photon will not be detected (and hence the pair will not be identified). That is due to the pixel dead time, rendering each pixel non-active for ~50 ns following each detection. The number of BX-X photon pairs missed in this manner can be estimated from the collected data by the following method. Given a BX-X pair detection, the probability the X photon will be detected in pixel $i$, $pX(i)_0$, is calculated by dividing the number of X detections in pixel $i$ by the overall BX-X pair number. $pBX(i)_0$, the probability to detect a BX photon in pixel $i$ given a pair detection, is calculated similarly. These probabilities are slightly biased due to the dead time described above, affecting pixels where higher intensity was measured more. We can estimate the probability for the BX and the X photons to impinge on the same pixel, $i$, as $pX(i)_0 \cdot pBX(i)_0$, and thus correct the probabilities in the following way:

$$pX(i)_1 = pX(i)_0 + pX(i)_0 \cdot pBX(i)_0 \tag{S4}$$

$$pBX(i)_1 = pBX(i)_0 + pX(i)_0 \cdot pBX(i)_0 \tag{S5}$$

The results shown in the manuscript are all after applying this correction.

# S5: Fitting to a dipole emission model for the X and BX

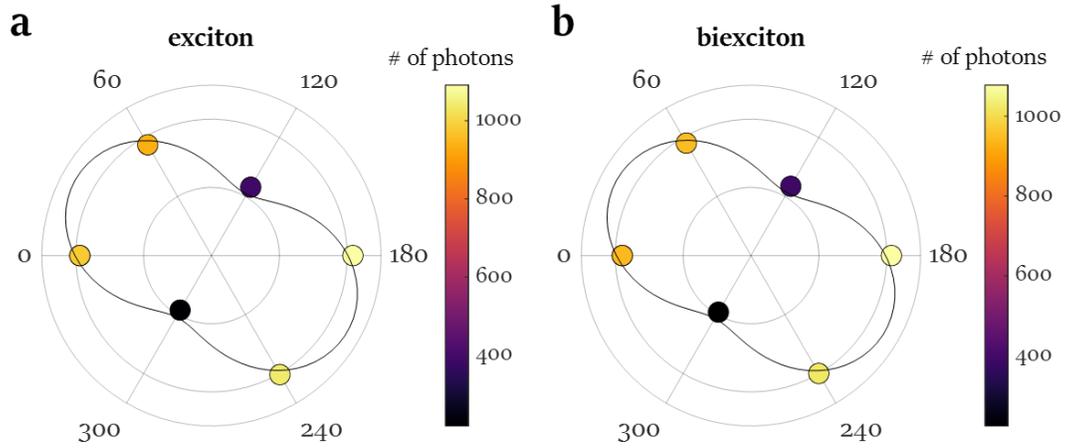

**Figure S4. Polar plot of the dipole fit for the exciton and biexciton.** Polar representation of the intensity values detected by the six pixels from the inner ring of the array (highlighted in figure 2b of the main text) for the exciton (a) and biexciton (b) data sets along with the fit (black solid line) to the integrated dipole emission model. The six data-points are colored according to the number of detected photons in each pixel. This is complementary to the measurement presented in figure 3e of the main text.

# S6: Heralded defocused imaging of single nanorods

## Single-particle analysis for NR1

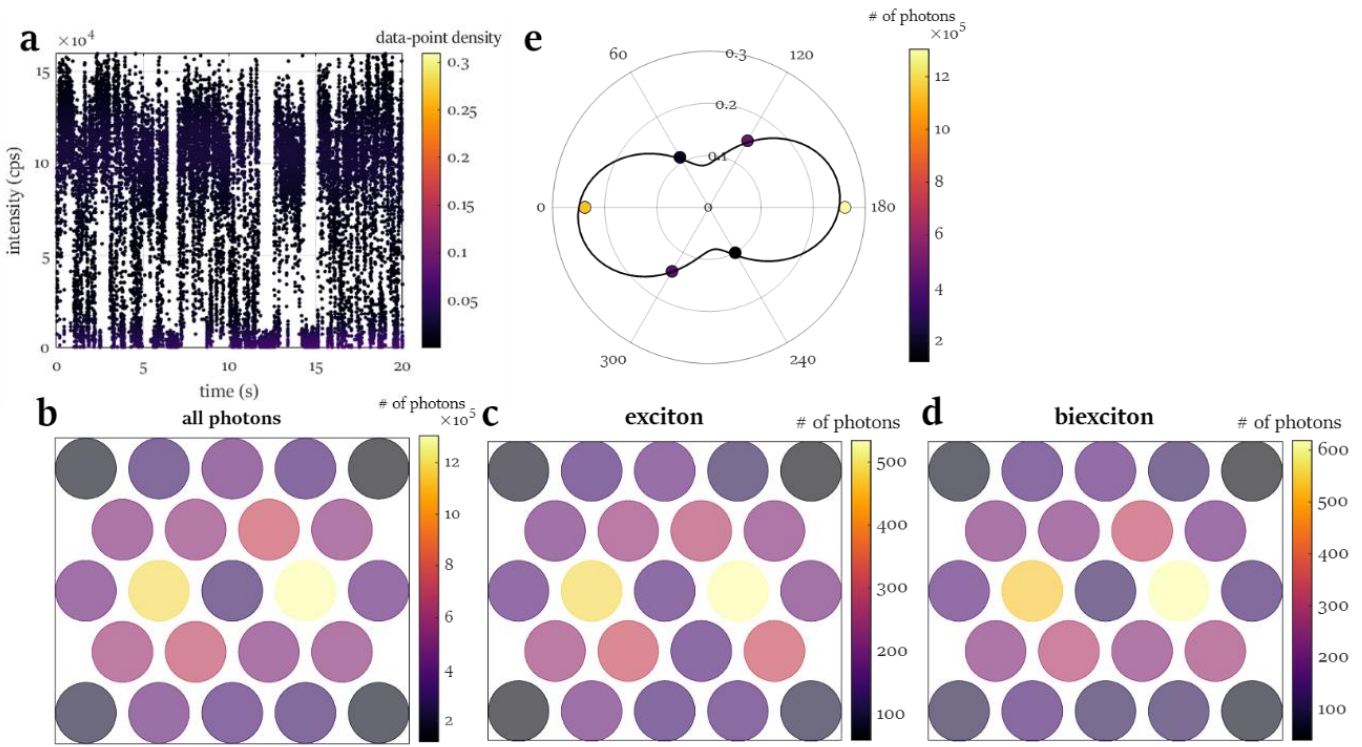

**Figure S5. Heralded defocused imaging analysis of a single NR1.** (a) Total fluorescence intensity collected for all pixels as a function of measurement time in a 20 second time window. (b-d) Histograms of all detected photons, titled "all photons" (b), and post-selected exciton (c) and biexciton (d) detections from a 5-minutes measurement of a single type-I½ CdSe/CdS seeded nanorod (NR1), applying heralded defocused imaging. Color scale represents the number of detections at a given detector pixel. (e) Polar representation of the intensity values detected by the six pixels from the inner ring of the array for the "all photons" data set along with the fit (black solid line) to a dipole emission model. The six data-points are colored according to the number of detected photons in each pixel.

## Single-particle analysis for NR2

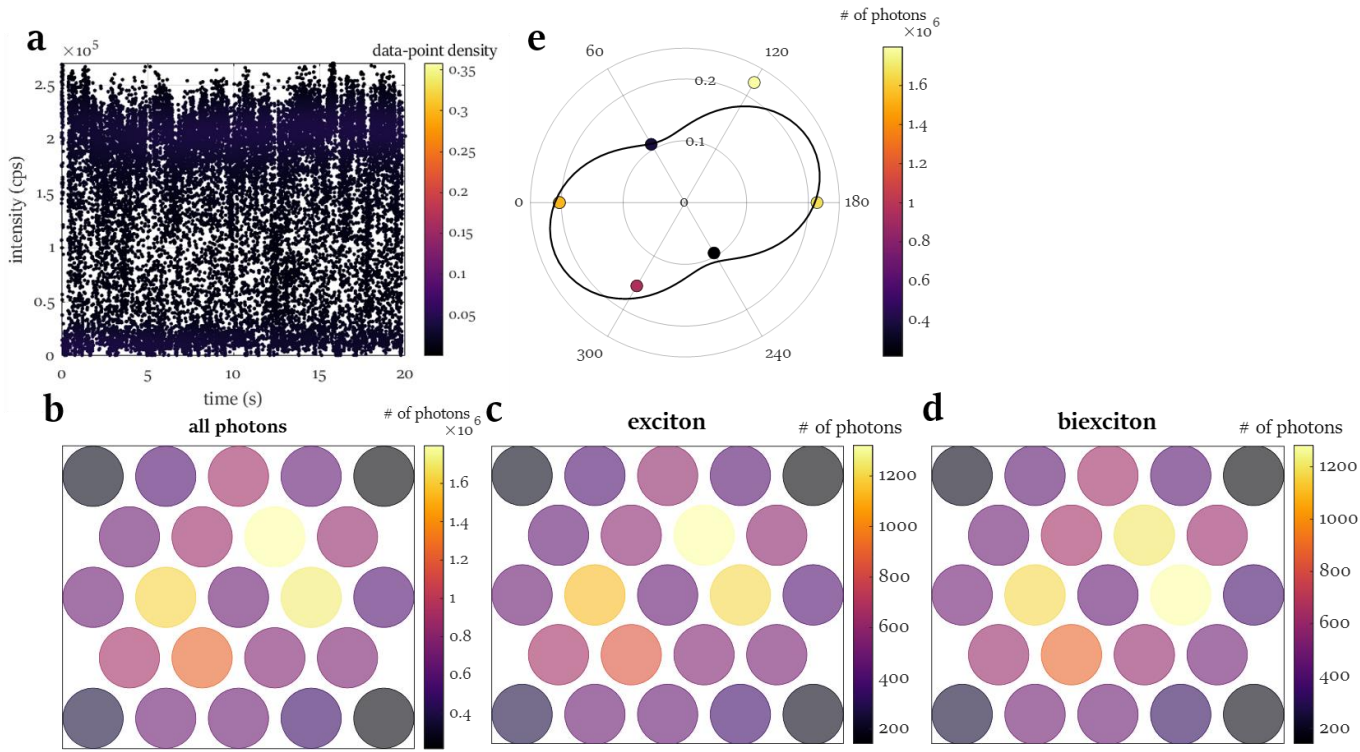

**Figure S6. Heralded defocused imaging analysis of a single NR2.** (a) Total fluorescence intensity collected for all pixels as a function of measurement time in a 20 second time window. (b-d) Histograms of all detected photons, titled "all photons" (b), and post-selected exciton (c) and biexciton (d) detections from a 5-minutes measurement of a single type-II ZnSe/CdS seeded nanorod (NR2), applying heralded defocused imaging. Color scale represents the number of detections at a given detector pixel. (e) Polar representation of the intensity values detected by the six pixels from the inner ring of the array for the "all photons" data set along with the fit (black solid line) to a dipole emission model. The six data-points are colored according to the number of detected photons in each pixel.

# S7: Correlation of the in-plane angle of the X and BX transition dipole moments

The fitting process described in section S5 results in both the anisotropy value of the X-to-GS and BX-to-X transition dipole moments of each NR and the in-plane angle of the dipoles. As seen in figure S7, the dipole angles of the X and BX are highly correlated, as expected, which strengthens and supports the anisotropy results.

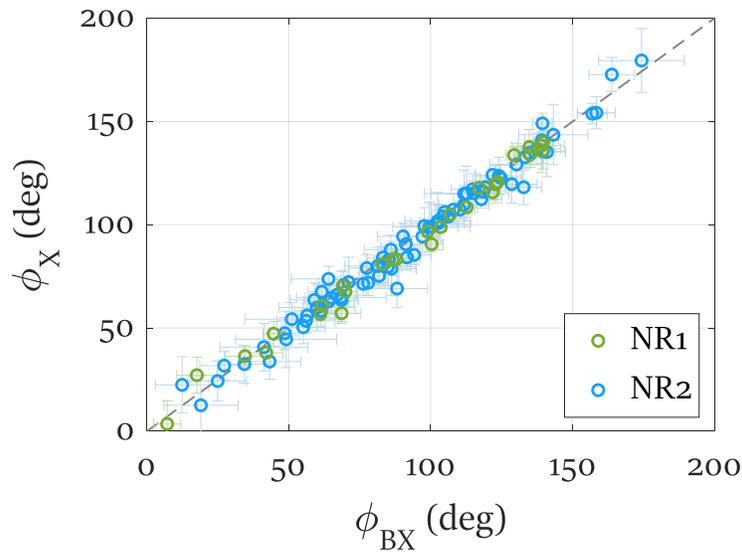

**Figure S7. In-plane angle correlation.** Correlation of the in-plane angle ($\phi$) of the X and BX transition dipole moments for 28 and 79 single measurements of NR1 and NR2, respectively.

## S8: Estimation of the dielectric effect

The following estimation of the dielectric effect in CdSe/CdS NRs (NR1) and ZnSe/CdS (NR2) was done according to the supporting information of ref 9. CdSe/CdS dot-in rods (similar to NR1 in this work) were approximated as homogeneous prolate ellipsoids with axes $a = b < c$, which feature anisotropic response to an excitation field. Table S1 presents the parameters needed for calculating the dielectric effect parameter, i.e. the ratio of the electric field strength between the major and minor axis: $R_e = \frac{f_c}{f_a}$, where $f_c$ and $f_a$ are called the local field factors. In our case $R_e > 1$, which implies that the attenuation of the electric field is stronger in the axes perpendicular to the rod's c-axis. Since the emission intensity scales with the square of the fields, we should take the square of $R_e$. $\Delta R_e$ represents the change in the dielectric effect of the CdS rod shell when the NR transitions from being doubly excited to singly excited, followed by a small refractive index change of ~0.4% (as mentioned in the main text). $\alpha_c$ and $\alpha_a$ are the depolarization factors of the major and minor axis respectively, and they are related to the geometry of the particles.[10]

Table S1. Calculation of the change in the dielectric effect parameter. $n_0$ is the steady state refractive index, $e$ is the eccentricity, $\alpha_c$ and $\alpha_a$ are the depolarization factors, and $R_e$ is the dielectric effect parameter.

|  | aspect ratio | $n_0$ of CdS at 600 nm | $e$ | $\alpha_c$ | $\alpha_a$ | $|\Delta R_e^2|/R_e^2$ |
|---|---|---|---|---|---|---|
| NR1 | 5 (5x25 nm) | 2.34 | 0.9798 | 0.0558 | 0.4721 | 0.96% |
| NR2 | 4 (8x32 nm) | 2.34 | 0.9682 | 0.0754 | 0.4623 | 0.88% |

## S9: Spectroscopy of short type-II ZnSe/CdS

An additional sample of short type-II ZnSe/CdS seeded NRs (with dimensions of 8x16 nm) was measured to examine NRs with smaller aspect ratio. Figures S8a,b show basic characterization of the NRs, with spectra similar to the long NR2 sample shown in figure 1c of the main text. The aggregate results demonstrated in figure S8c present 23 single measurements. The diagonal is a guide to the eye, representing equal anisotropy values for the X and BX. In this correlation, all data points are distributed around the diagonal almost equally. The histogram of $\Delta_{anisotropy} = X_{anisotropy} - BX_{anisotropy}$ is demonstrated in figure S8d. Interestingly, the positive mean value suggests a trend similar to the long type-II NRs (NR2) shown in figure 4 of the main text, in which the X-to-GS transition dipole is more anisotropic than the BX-to-X transition dipole. Yet, this trend is much less significant due to lack of statistics. The statistical significance was estimated by a paired Student's t-test and yielded a score of 1 (corresponding to a p-value of ~0.3).

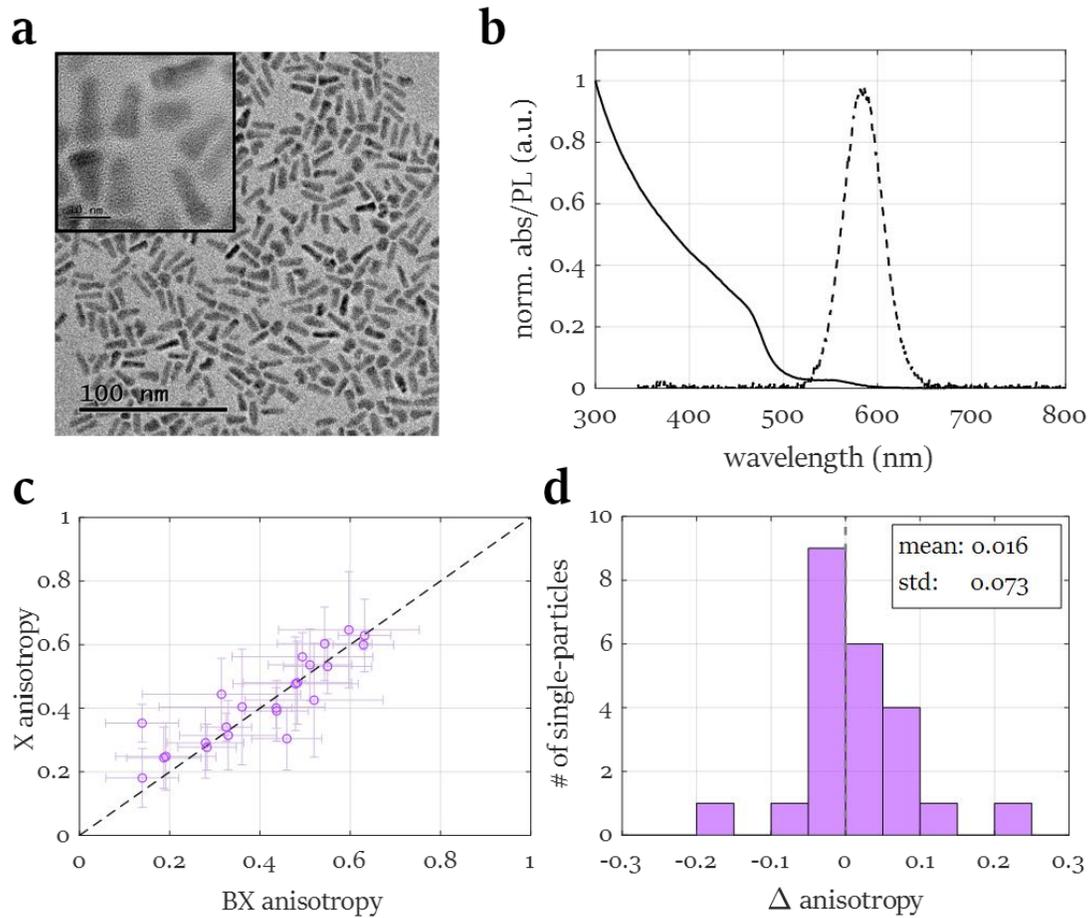

**Figure S8. Characterization and heralded defocused imaging results for short type-II ZnSe/CdS seeded nanorods (NRs).** (a) Transmission electron microscope image of the NRs. Inset is a high-resolution image. Scale bar is 10 nm. (b) Normalized absorbance and photoluminescence of the NRs. Emission peak is at ~583 nm. (c) Exciton anisotropy versus biexciton anisotropy values extracted from the fit. (d) Histogram of delta anisotropy values of the NRs. The dashed grey line in c and d is a guide to the eye, indicating the same anisotropy values for both photons.